\documentclass[smallextended]{svjour3}       
\smartqed  
\usepackage{graphicx}
\usepackage{amsmath}
\usepackage{amsfonts}
\usepackage{lineno}
\usepackage{array}
\usepackage{longtable}
\usepackage{color}
\usepackage[normalem]{ulem}
\usepackage{hyperref}

\begin{document}

\title{Uncertainty Relations for Mesoscopic Coherent Light
}


\author{Ariane Soret$^{1,2,3,4}$ \and  Ohad Shpielberg$^{5,6,*}$ \and Eric Akkermans$^4$ 
}

\institute{$^1$ \at Complex Systems and Statistical Mechanics, Department of Physics and Materials Science, University of Luxembourg, L-1511 Luxembourg, Luxembourg \and
$^2$ \at Laboratoire Matériaux et Phénomènes Quantiques, Université Paris Diderot, Sorbonne Paris Cité and CNRS, UMR 7162, F-75205 Paris Cedex 13, France \and
$^3$ \at CPHT, Ecole Polytechnique, CNRS, Universit\'e Paris-Saclay, Route de Saclay, 91128 Palaiseau, France \and
$^4$ \at Department of Physics, Technion -- Israel Institute of Technology, Haifa 3200003, Israel \and
$^5$ \at Haifa Center for Theoretical Physics and Astrophysics, Faculty of Natural Sciences, University of Haifa, Haifa 3498838, Israel \and
$^6$ \at Collège de France, 11 place Marcelin Berthelot, 75231 Paris Cedex 05 - France,  \\$^*$\email{ohads@sci.haifa.ac.il}}

\date{Received: DD Month YEAR / Accepted: DD Month YEAR}

\maketitle

\begin{abstract}

Thermodynamic uncertainty relations unveil useful connections between  fluctuations in thermal systems and entropy production. This work extends these ideas to the disparate field of \textit{zero temperature} quantum mesoscopic physics where fluctuations are due to coherent effects and entropy production is replaced by a cost function. The cost function arises naturally as a bound on fluctuations, induced by coherent effects -- a critical resource in quantum mesoscopic physics. Identifying the cost function as an important quantity demonstrates the potential of importing powerful methods from non-equilibrium statistical physics to quantum mesoscopics.  


\end{abstract}

\section{Introduction}
\label{intro}

The study of non-equilibrium physics led to a wealth of useful results, e.g. linear response, fluctuation theorems, Onsager relations, exact models and effective hydrodynamic descriptions \cite{LifshitzStatPhys,Derrida07}. These approaches are implemented in the realm of systems where the underlying stochastic nature results mainly from thermal noise. It is known that a system at thermal equilibrium fluctuates and the probability of rare but significant fluctuations are estimated from the Einstein formula. Although non-equilibrium physics requires new approaches different from the familiar thermodynamics concepts, it is intuitively helpful to relate these two situations. Le Chatelier principle states that at thermodynamic equilibrium, the net outcome of a fluctuation is to bring the system back to equilibrium namely, thermodynamic potentials are either concave or convex functions. The Onsager formulation allows to extend Le Chatelier principle to non-equilibrium physics. A system brought out of equilibrium by the application of forces $X_\alpha$, such as temperature or density gradients, creates currents $j_\alpha$ linearly related to the forces, $j_\alpha = \sum_\beta L_{\alpha \beta} \, X_\beta $ such that products $j_\alpha \, X_\alpha$ are additive terms in the entropy creation $\dot{\Sigma}_{th} = \sum_\alpha j_\alpha \, X_\alpha$ per unit time. The symmetric matrix $L_{\alpha \beta}$ cannot be determined from thermodynamics but only from a microscopic model. 

Useful and simple inequalities, $\mathcal{F}_\alpha \langle \Sigma_{ th} \rangle_{th} \geq 2 $ between the entropy production and fluctuating currents $j_\alpha$ have been obtained recently \cite{Barato15} where  $\mathcal{F}_\alpha \equiv (\langle j_\alpha ^2 \rangle_{th} - \langle j_\alpha  \rangle_{th} ^2) / \langle j_\alpha  \rangle_{th} ^2$  and  $\langle\cdot\cdot\cdot\rangle_{th} $ is an appropriate thermal average. These inequalities, termed thermodynamic uncertainty relations (TUR), have triggered significant effort \cite{Barato15,Gingrich2016,HorowitzReview} exploring their generality \cite{Koyuk2019,Timpanaro2019,Sasa18,Niggemann2020,Kashia2018,Falasco20} and the universal, i.e.\ independent of specific details, lower bound. They  provide quantitative criteria to evaluate the tradeoff between fluctuations and their cost, so as to produce currents with a certain precision.  
TUR were successfully applied to assess energy input required to operate a clock or bounding the number of steps in an enzymatic cycle \cite{Barato15,Seifert18}, and deriving the efficiency of  molecular motors  \cite{Pietzonka2016}.  Finally, TUR inspired further studies  of entropy production bounds under certain constraints \cite{Raz2016,Busiello2018,Shpielberg2018}. 

The purpose of this paper is to present a novel and non anticipated approach to benchmark TUR underlying ideas and to check them in physical setups easily accessible experimentally. Concretely, we consider the problem of propagation of quantum or classical waves in random media. A wealth of measurable features about this problem has been  achieved using so called incoherent approximations, namely washing out interferences between waves. Yet, in certain limits, remaining interference effects are observable and at the basis of spectacular and measurable phenomena, e.g. weak and strong localisation \cite{Anderson57}, generally known as quantum mesoscopic effects \footnote{The denomination "quantum mesoscopic physics" that we shall keep, may seem to indicate that such remaining interferences occur only in quantum systems of intermediate sizes in between macroscopic and microscopic. It is not necessarily so. This name has been coined historically after identifying these interferences as quantum effects in conductors of mesoscopic sizes so as to minimise incoherent and inelastic processes.} \cite{Akkermans}. 

We wish to establish a correspondence between these  effects and fluctuating non-equilibrium systems, where fluctuations induced by coherent effects play the role of thermal fluctuations. This correspondence makes mesoscopic coherence induced fluctuations eligible on their own to a non thermal kind of uncertainty relations, henceforth coined quantum mesoscopic uncertainty relations (QMUR).

We also define a cost function $\Sigma$, analogous to the entropy production $\Sigma_{\rm th}$, so as to set a lower bound and a trade-off for phase coherent induced fluctuations $\langle f^2 \rangle$ for  relevant mesoscopic quantities $f$, namely, 
\begin{equation}
    \langle f^2 \rangle_c \, \langle \Sigma \rangle \geq 2 \langle f \rangle^2
    \label{qmur}
\end{equation}
where $ \langle f^2 \rangle_c =  \langle f^2 \rangle -  \langle f \rangle^2$ and $\langle \cdots \rangle$ denotes an average over disorder realizations.

The mapping we propose between quantum mesoscopic and non-equilibrium physics appears to be beneficial to both fields. It suggests an alternative benchmark approach to non-equilibrium physics features, e.g., entropy production rate, large deviation functions, thermal uncertainty relations (TUR), Fisher information \cite{seifert2012stochastic} and fluctuation induced forces \cite{Kardar99}. Conversely, by  importing novel tools and concepts from non-equilibrium physics to quantum mesoscopics, this mapping  allows to address pending issues in this thoroughly studied field, e.g. new types of control to the strength and feasibility of mesoscopic coherent effects,  but also to propose new measurable physical quantities such as cost function and long range mechanical forces induced by coherent fluctuations.

\section{Outline}
\label{sec:1}

\subsection{Scope}
The scope of this paper is to show that coherent effects in the  propagation of waves in random media can be quantitatively described using an approach akin to thermal non-equilibrium systems, where fluctuations induced by coherent effects play the role of thermal fluctuations and lead to uncertainty relations. To establish this new kind of uncertainty relations (QMUR), we define a cost function $\Sigma$, analogous to the entropy production  $\Sigma_{\rm th}$. Then, we establish an expression for the average cost function $\langle \Sigma \rangle$ and apply it to show the genuine interest of QMUR to optimise quantum mesoscopic features in different setups.  

\subsection{Structure of the paper}

The paper is organised as follows. In section 3, we  introduce in layman's terms ideas underlying coherent effects in quantum mesoscopic physics. In section 3.1, we present basic material on the well accepted diffusive limit for the spatial behavior of the incoherent intensity of the wave field. In section 3.2, we discuss how the microscopic time reversal symmetry  lost at the level of the diffusive and incoherent wave propagation is restored perturbatively in the weak scattering limit. This essential idea that interference effects are related to time reversal invariance is further discussed in section 3.4 in the equivalent language of a stochastic Langevin equation where the noise is solely driven by spatially local interference terms.

Section 4 is devoted to a phenomenological description of quantum mesoscopic uncertainty relations (QMUR) using Onsager description so as to provide some physical intuition about their meaning. In section 5, we derive QMUR in the more general framework of statistical field theory. This allows to define a cost function at the trajectory level. A generalised form of QMUR is given in section 5.3. Section 6 contains examples to illustrate the meaning and calculation of QMUR in the geometry of a slab and for fluctuating forces. An alternative derivation of QMUR is presented in section 7 which is based on the Cramer-Rao bound hence unveiling a relation with Fisher information. In section 8 we conclude and discuss further potential applications of our results.  


\section{Quantum Mesoscopic Physics (QMP)}

Quantum mesoscopic physics is devoted to study  waves (quantum or classical) propagating in a random potential. To maintain a homogeneous description throughout the paper, we opt for the language of propagation of scalar waves and consider a random and $d$-dimensional dielectric medium of volume $V = L^d$ illuminated by a monochromatic scalar radiation of wave-number $k$ incident along a direction of unit vector $\mathbf{\widehat{k}}$. Inside the medium, the amplitude $E(\mathbf{r})$ of the radiation is solution of the scalar Helmholtz equation,
\begin{equation}
\Delta E(\mathbf{r})+k^2\left(1+\mu(\mathbf{r})\right)E(\mathbf{r})=s_0(\mathbf{r})
\label{helmoltz}
\end{equation}
where $\mu(\mathbf{r})=\delta \epsilon(\mathbf{r})/\langle\epsilon\rangle$ denotes the fluctuation of the dielectric constant $\epsilon(\mathbf{r}) = \langle\epsilon\rangle+ \delta \epsilon(\mathbf{r})$, $\langle\cdot\cdot\cdot\rangle$ is the average over disorder realizations and $s_0(\mathbf{r})$ is the source of radiation. Obtaining solutions of the Helmholtz equation in the presence of a random potential is an arduous task, namely despite its simple formulation, this problem is notoriously rich and difficult. A popular and useful approach starts from a description of the temporal evolution of a wave packet in random media \cite{Akkermans}. This method uses the formalism of Green’s functions known to facilitate an iterative expansion in powers of the disorder potential, also called multiple scattering expansion. In the limit of weak disorder, which we will define properly, this expansion is expressed in the form of a series of independent processes, termed collision events, separated by a characteristic time $\tau$, the elastic collision time evaluated using the Fermi golden rule. Associated to $\tau$ is a characteristic length, the elastic mean free path $l$, defined by $l = v \tau$, where $v$ is a conveniently defined group velocity of the wave.  Together with the wave-number $k$, the radiation in the random medium is thus characterized by two length scales, $k^{-1}$ and $l$. Equipped with the multiple scattering expansion, it is possible to calculate relevant disorder averaged physical quantities in perturbation in the so-called weak disorder/scattering limit $\left( k l \gg 1 \right)$. For the rest of the paper, we consider the three dimensional case $d=3$.

\subsection{Diffusion Equation}
\label{sec:Diffusion Equation}

The multiple scattering expansion advocated in the previous section allows to describe the evolution of a plane wave in a random medium, i.e. technically to evaluate the disorder averaged (one-particle) Green's function of (\ref{helmoltz}).  But it does not contain information about the spatial and time evolution of a wave packet. For optically thick media, most physical properties are determined not by the average Green's function, but rather by the two-particle or intensity Green's function $P( \mathbf{r},\mathbf{r'})$, associated to the behaviour of the radiation intensity $I(\mathbf{r}) = |E(\mathbf{r})|^2$.

A convenient way to illustrate these ideas \cite{Akkermans} is to start from the expression of the one-particle Green's function,
\begin{equation}
G (\mathbf{r}, \mathbf{r}',k) = \sum_{N=1}^{\infty}
\sum_{\mathbf{r}_1,\cdots,\mathbf{r}_N}|A (\mathbf{r}, \mathbf{r}', {\cal C}_N)| \exp ( i  k {\cal
L}_N ) \label{greentraj}
\end{equation}
where $A (\mathbf{r}, \mathbf{r}', {\cal C}_N) $ is the complex valued amplitude associated to a multiple scattering sequence of length 
 $N$, ${\cal C}_N =
(\mathbf{r}_1, \mathbf{r}_2,..., \mathbf{r}_N)$. The accumulated phase $k {\cal
L}_N $ measures the length $ {\cal L}_N $ of the multiple scattering sequence in units of the wavelength $ k^{-1}$. 

The two-particle Green's function $P( \mathbf{r},\mathbf{r'})$ is proportional to $G G^*$, 
hence it involves an accumulated phase given by the length difference ${\cal L}_N - {\cal L}_{N'}$ between any two multiple scattering sequences. Upon averaging over disorder, it can be anticipated that only identical multiple scattering sequences ${\cal L}_N = {\cal L}_{N'}$ up to a single scattering event $l$, will survive the large phase scrambling in the weak scattering limit $k l \gg 1$. Keeping only these two coupled identical one-particle Green's functions trajectories leads to an approximate  two-particle phase independent intensity Green's function $P_D( \mathbf{r},\mathbf{r'})$ as represented in Fig.~\ref{fig-dis-rev}.a.

Building on this result known as the incoherent limit, a wealth of phenomenological descriptions has been proposed. Among them, radiative transfer describes the disorder average macroscopic wave intensity $I_D(\mathbf{r})$ and the associated current $\mathbf{j}_D$, obtained by keeping only incoherent, i.e.\ phase independent, contributions  \cite{Akkermans,Ishimaru}. They are related by a Fick's law, $\mathbf{j}_D(r) = -D\boldsymbol{\nabla} I_D(\mathbf{r})$ with $D = vl/3$ being the diffusion coefficient, which together with current conservation $\nabla \cdot \mathbf{j}_D =0$, leads to a steady state diffusion equation, $-D\Delta I_D(\mathbf{r}) = 0$ with boundary conditions ensuring the vanishing of the incoming diffusive flux (see appendix \ref{app:BC radiative transfer}).

The main drawback of these phenomenological approaches is their neglecting of interference effects washed out by the disorder averaging. Yet, phase coherence  is not erased by the disorder average and is at the origin of spectacular measurable effects, e.g.\ Anderson localization (weak and strong), coherent backscattering, universal conductance fluctuations and Sharvin $\&$ Sharvin effect to cite a few (a selection of the extremely vast literature on these topics is accessible in e.g.\ \cite{Akkermans}).

It is worthwhile discussing the role of time reversal symmetry (TRS) in these interference effects. At the level of the wave equation (\ref{helmoltz}), the multiple scattering amplitude and hence the one-particle Green's function (\ref{greentraj}) are invariant under time reversal \footnote{The notion of time reversal symmetry is sometimes presented as equivalent to reciprocity. It is not so, e.g. in the presence of absorption, an irreversible process, time reversal symmetry is broken but not reciprocity.}. Namely, $G(\mathbf{r},\mathbf{r}',\mathbf{k}) = G^* (\mathbf{r}' , \mathbf{r}, -\mathbf{k})$ \cite{Akkermans}. TRS is broken in the incoherent diffusive approximation. To see it, note from Fig.~\ref{fig-dis-rev}.a, that TRS can be implemented either by reversing the two conjugated multiple scattering sequences in the two-particle Green's function $P_D( \mathbf{r},\mathbf{r'})$ or by reversing only one, leaving the second sequence unchanged. In the latter case, the resulting two-particle Green's function cannot be written as an incoherent function $P_D( \mathbf{r},\mathbf{r'})$ hence TRS is broken in the incoherent diffusive limit. In the weak scattering limit $k l \gg 1$, it has been shown that reversing a single sequence leads to a two-particle Green's function given by $P_D( \mathbf{r},\mathbf{r'})$ times a small and local correction known as a quantum crossing \cite{Akkermans}. Therefore, quantum crossings are a signature of a broken TRS. 

These results can be made more systematic using a semi-classical description which enables to include coherent effects in the incoherent radiative transfer model.

\subsection{Quantum Crossings}

\begin{figure}[tb]
    \centering
    \includegraphics[scale=0.40]{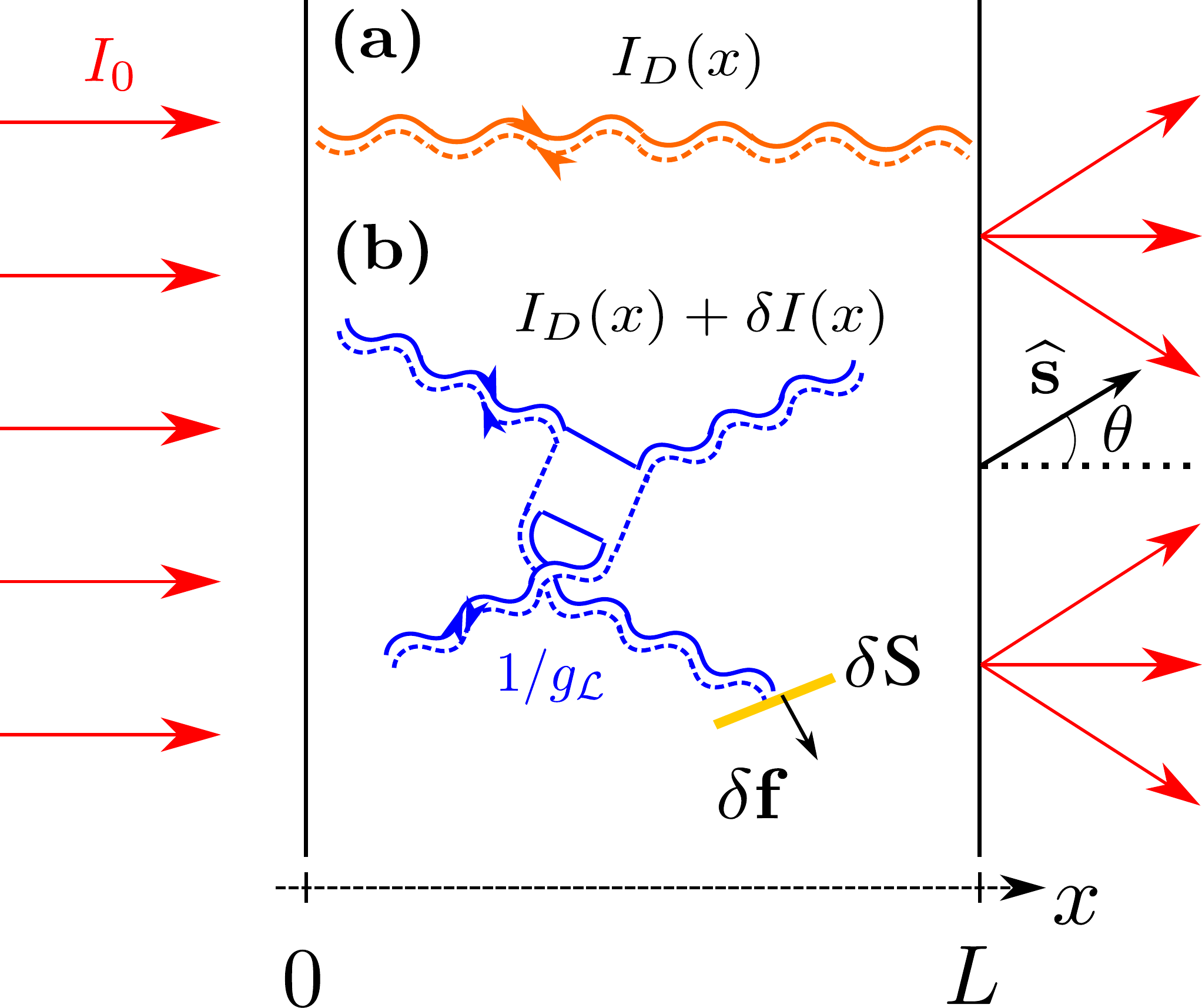}
    \caption{A slab of section $S$ and width $L$ is filled with a scattering medium, and is illuminated by a monochromatic plane wave. In this geometry, $\mathcal{L}=S/L$, hence $g_\mathcal{L} = k^2 l S / 3 \pi L$. {\bf (a)} The diffusive intensity, $I_D(x)=P_D(0,x)$,  
    is built out of paired multiple scattering amplitudes represented by full and doted wave shaped lines. 
    {\bf (b)} Coherent intensity fluctuations $\delta I (x)$ are described by schematically represented quantum crossings, accounted by the noise (\ref{noise-ampl}). Exchange of multiple scattering amplitudes and new pairings occur within a small ($\propto 1/g_\mathcal{L} $) volume, and induce a small dephasing. Intensity fluctuations induced by quantum crossings have been observed, e.g by measuring light transmitted along a direction $\mathbf{\widehat{s}}$ \cite{Scheffold98,Scheffold97} and could be measured by means of the predicted fluctuations $\delta\mathbf{f}$ of the radiative force exerted on a suspended membrane of surface $\delta S$ (yellow in the figure) \cite{Soret19}. 
    \label{fig-dis-rev}
    } 
\end{figure}

The semi-classical approach starts from the formal sum (\ref{greentraj}) over multiple scattering trajectories. As already stated, each phase independent, incoherent  trajectory obtained for the diffusive intensity $I_D(\mathbf{r}) \propto G(\mathbf{r_0,r} )G^*(\mathbf{r,r_0})$, where $\mathbf{r_0}$ is the location of the light source\footnote{This holds for a point source located at $\mathbf{r_0}$. For an extended light source, $I_D$ is obtained by performing an integral over $\mathbf{r_0}$.}, is built from the pairing of two identical, but complex conjugated, multiple scattering amplitudes $G$ and $G^*$. By construction, these two amplitudes have opposite phases so that the resulting diffusive trajectory is phase independent (Fig.~\ref{fig-dis-rev}.a). Unpairing these two sequences gives access to the underlying phase $k {\cal
L}_N $ carried by each multiple scattering amplitude and thereby to phase dependent corrections. The semi-classical description makes profit of this remark to evaluate systematically phase coherent corrections which correspond to a local crossing \cite{Hikami81}, where two diffusive trajectories mutually exchange their phase so as to form two new phase independent diffusive trajectories (Fig.~\ref{fig-dis-rev}.b). This local crossing -- a.k.a quantum crossing -- 
irrespective to the exact nature of waves, is a phase-dependent correction propagated over long distances by means of diffusive incoherent trajectories. Yet, the exact local structure of a quantum crossing depends on the exact nature of the wave, its degrees of freedom and applied fields. This picture for coherent mesoscopic effects is presented at an introductory level in \cite{Akkermans} (section 1.7). The occurrence of a quantum crossing  is controlled by a single dimensionless parameter $g_{\mathcal{L}}$ known as the conductance which depends on scattering properties and on the geometry of the medium. For a three dimensional ($d=3$) setup, the conductance $g_{\mathcal{L}}$ is
\begin{equation}
g_{\mathcal{L}} \equiv \frac{k^2 l}{3\pi}\mathcal{L} \, 
\label{g}
\end{equation}
where the macroscopic length $\mathcal{L} (\gg l)$ depends only on the geometry.
 In the weak scattering limit $k l \gg 1$, the conductance $g_{\mathcal{L}} \gg 1$ and small coherent corrections  generated by quantum crossings show up as powers of $1 / g_{\mathcal{L}}$. This scheme allows to expand relevant physical quantities, e.g.\ spatial correlations of the fluctuating intensity $\delta I (\mathbf{r}) \equiv I (\mathbf{r}) - I_D(\mathbf{r})$ as
\begin{equation}
\frac{\langle \delta I (\mathbf{r}) \delta I (\mathbf{r'})\rangle}{I_D(\mathbf{r})I_D(\mathbf{r'})} = C_1(\mathbf{r,r'})  + C_2(\mathbf{r,r'})  + C_3(\mathbf{r,r'}) 
\label{correlations}
\end{equation}
where the first contribution $C_1(\mathbf{r,r'}) = \frac{2\pi l}{k^2}\delta(\mathbf{r-r'})$ is short ranged and independent of $g_{\mathcal{L}}$. The two other contributions are long ranged, and respectively proportional to $1/g_{\mathcal{L}}$ and $1/ g_{\mathcal{L}}^2$. All three terms contribute to specific features of long ranged interference speckle patterns, and have been measured in weakly disordered electronic and photonic media \cite{Scheffold98,Scheffold97}.

\subsection{Effective Langevin Equation}
\label{subsec: effective langevin equation}

Essentially, all previous considerations stem from the remark that  spatially long ranged coherent effects result from short range phase-dependent quantum crossings occurring at scales $\ll l$ (see Fig.~\ref{fig-dis-rev}.b). 
Stated otherwise, the large scale coarse grained hydrodynamic description of incoherent light can be modified to include coherent effects by inserting a local, properly tailored, noise function so as to reproduce expected long range coherent effects. Building on this remark, an elegant and systematic description has been proposed \cite{Spivak87}, based on the Langevin equation, 
\begin{equation}
\mathbf{j(\mathbf{r})}=-D \boldsymbol{\nabla} I(\mathbf{r})+  \boldsymbol{\xi}\mathbf{(r)} \, 
\label{langevin}
\end{equation}
for the mesoscopic quantities $I (\mathbf{r})$ and $\mathbf{j(r)}$. Disorder averaging is performed only at large scales $\geq l$ hence the stochastic nature of both quantities. 
This stochastic approach while phenomenological in nature, is equivalent to a perturbation theory for microscopic quantities with respect to the small and dimensionless parameter $1 / g_{\mathcal{L}}$.
The time-independent noise $\boldsymbol{\xi}\mathbf{(r)}$ includes all the information relative to phase coherence induced by quantum crossings. Its spatial correlations are systematically calculable as powers of $1 / g_{\mathcal{L}}$. The $g_{\mathcal{L}}$-independent behaviour accounts for the incoherent diffusive limit. The details of this generally cumbersome but well understood procedure are presented in \cite{Soret19} \cite{ThesisSoret19}. 
The noise has zero mean and 
to lowest order $1 / g_{\mathcal{L}}$ it is Gaussian \cite{Soret19,Spivak87}, 
\begin{equation} \langle\xi_{\alpha}(\mathbf{\mathbf{r}})\xi_{\beta}(\mathbf{\mathbf{r'}})\rangle =\sigma(I_D) \delta_{\alpha\beta}\, \delta(\mathbf{r-r'})\label{noise}
\end{equation} 
with the conductivity 
\begin{equation}
\sigma(I) = \frac{2\pi l v^2}{3k^2} \, I^2 (\mathbf{r}) 
\label{noise-ampl}
\end{equation}
similar to thermal diffusive processes \cite{Shpielberg2016}. Note that, to lowest order  $1/g_\mathcal{L}$, \eqref{langevin} is a weak noise Langevin equation. Namely, 
relative to the mean current, $\boldsymbol{\xi}$ scales like $1/\sqrt{g_{\mathcal{L}}} \ll 1$ (see Appendix \ref{app: dimensionless scaling of Langevin}). 
The Langevin equation (\ref{langevin}) based on the two parameters $D$ and  $\sigma$, provides a complete hydrodynamic description of the coherent light flow in a random medium and it extends Fick's law to the fluctuating mesoscopic quantities $I (\mathbf{r}) \equiv I_D (\mathbf{r}) + \delta I (\mathbf{r})$ and $\mathbf{j(r)} \equiv \mathbf{j_D(r)} + \delta \mathbf{j(r)}$.

It is important to emphasize that the noise $\boldsymbol{\xi}\mathbf{(r)} $ accounts for  phase-dependent  corrections (quantum crossings) and not for the random disorder in (\ref{helmoltz}). Note also that $\boldsymbol{\xi}\mathbf{(r)} $ does not restore TRS.


The form (\ref{noise-ampl}) of the noise is appealing since a constant $D$ and $\sigma(I) \propto I^2$, correspond to the Kipnis-Marchioro-Presutti (KMP) process -- a heat transfer model for boundary driven one dimensional chains of mechanically uncoupled oscillators strongly out of equilibrium \cite{Bertini05,Shpielberg2016,Kipnis82}, well described by the macroscopic fluctuation theory \cite{Bertini15,Jordan}. Hence, the Langevin equation (\ref{langevin}) driven by local coherent processes (\ref{noise}) suggests to deepen the analogy with thermal diffusive non-equilibrium steady states. To that aim, we recall in the next section some basic concepts in thermal non-equilibrium physics so as to define a cost function and prove a new type of uncertainty relations (QMUR).

\section{Onsager description and QMUR }


The statistical interpretation of the entropy by Boltzmann and Einstein is at the heart of statistical mechanics as well as modern application to non-equilibrium statistical mechanics in the form of large deviations \cite{LDF_Boltzmann}. Following this success, it is not surprising that entropy-like descriptions have been proposed for athermal systems like jammed granular matter \cite{Baule_Review,DeGiuli_Edwards}  and more recently for data compression \cite{Levine_Data}. 

Entropy production -- a measure on how far a system is from equilibrium --  has a central role in the study of relaxation to equilibrium, dissipation in non-equilibrium steady states and in the efficiency of thermal engines (see \cite{seifert2012stochastic} and references within).  Despite its importance, as far as we know, no attempts were made to extend the definition of entropy production to athermal non-equilibrium systems.   

The purpose of this section is to suggest an expression to the cost function -- the analog of entropy production in quantum mesoscopics. First, we recall how to obtain entropy production for thermal, non-equilibrium steady state  and in particular for thermal diffusive systems. Then, we take advantage of the analogy between thermal diffusive systems and effective Langevin description \eqref{langevin} for quantum mesoscopics to define our cost function. We conclude by proving the QMUR \eqref{qmur}. A physical interpretation of the cost function is postponed to section~\ref{sec:Stochastic integral}. To avoid confusion between thermal and mesoscopic quantities, when relevant we add the subscript $(\cdots)_{th}$ for thermal.

\subsection{Entropy production in thermal non-equilibrium steady states}

Consider a thermal system, coupled to two reservoirs keeping it in a non-equilibrium steady state and sustaining an energy and particle steady state currents $\mathbf{J}_u , \mathbf{J}_\rho$. We assume that the macroscopic system can be divided into small systems still macroscopic in nature, but that are slightly out of equilibrium. Hence, the entropy density for each subsystem is $ds = \frac{1}{T}du - \frac{\mu_c}{T} d\rho $ where $T,\mu_c$ are the local temperature  and chemical potential, and $u,\rho$ are the energy and particle densities. We further assume that energy and density are locally conserved; $\dot{\rho}  + \nabla \cdot \mathbf{J}_\rho  = 0 $   and $\dot{u}  + \nabla \cdot \mathbf{J}_u= 0 $. The entropy flux is thus $\mathbf{J}_s = \frac{1}{T} \mathbf{J}_u  -\frac{\mu_c}{T} \mathbf{J}_\rho  $. The steady state entropy production rate $\langle \dot{\Sigma}_{th} \rangle_{th}$  in each subsystem is defined as the excess from the conservation equation, i.e. $\langle \dot{\Sigma}_{th} \rangle_{th} = (\dot{s} + \nabla \cdot \mathbf{J}_s)d \mathbf{r} $. This implies that for the entire system  $\langle \dot{\Sigma}_{th} \rangle_{th} = \int d \mathbf{r} \, (\nabla \frac{1}{T})\cdot \mathbf{J}_u + (\nabla \frac{-\mu_c}{T})\cdot \mathbf{J}_\rho $. This result can be generalized to account for other thermodynamic forces.  

\subsection{Application to thermal diffusive systems}
\label{sec:thermal diffusion}

We focus now on thermal non-equilibrium steady state diffusive systems at uniform temperature here set to unity. The steady state current is then given by Fick's law $\mathbf{J}_\rho = - D_{th} \nabla \langle \rho \rangle_{th}  $, where $D_{th}$ is the corresponding diffusion coefficient and $\langle \rho \rangle_{th} $ is the steady state density profile. The fluctuating hydrodynamics \cite{Bertini15,Lecomte_eqlikefluc}  describes the fluctuations of the diffusive dynamics 
\begin{eqnarray}
\label{eq:fluc hyrdo}
\partial_t \rho &=& - \nabla \cdot \mathbf{j}_\rho
\\ \nonumber 
\mathbf{j}_{\rho} &=& -D_{th} \nabla \rho + \eta, 
\\ \nonumber 
\langle \eta_{\alpha}  (\mathbf{r},t) \eta_{\beta} (\mathbf{r}',t') \rangle_{th} &=& \delta_{\alpha \beta} \delta(\mathbf{r}-\mathbf{r}')\delta(t-t')\sigma_{th}(\rho)  
\end{eqnarray}
where $\langle \eta \rangle_{th}$ vanishes. Note the analogy between time-independent fluctuations in \eqref{eq:fluc hyrdo} and the mesoscopic Langevin equation \eqref{langevin}.

The conductivity $\sigma_{th}$ and diffusion $D_{th}$ are not independent and abide the Einstein relation $f''(\rho)  = 2D_{th}/\sigma_{th}$, where $f(\rho)$ is the free energy density. Moreover, since  $\mu = f'(\rho)$, previous considerations allow to identify the entropy production in the thermal system as\footnote{$D_{th},\sigma_{th}$ are evaluated at $\langle \rho \rangle_{th}$. }
\begin{equation}
\label{eq:thermal EP}
    \langle \Sigma _{th} \rangle_{th} = \int dt \, \langle \dot{\Sigma} _{th} \rangle_{th} = \int dt  \int d \mathbf{r} \frac{2D^2 _{th} }{\sigma_{th}} (\nabla \langle \rho \rangle_{th} )^2.
\end{equation}
The analogy just mentioned between thermal diffusive fluctuations \eqref{eq:fluc hyrdo} and quantum mesoscopic fluctuations \eqref{langevin},
 allows to define a cost function $\langle \Sigma \rangle $ for coherent diffusive quantum mesoscopic systems.

\subsection{Proof of QMUR using the cost function}

Based on \eqref{eq:thermal EP} and \eqref{langevin}, we propose for the disorder averaged cost function $\langle \Sigma \rangle$, the phenomenological expression, 
\begin{equation}
\label{eq: mean cost def}
    \langle \Sigma \rangle  = \int d \mathbf{r} \frac{2D^2}{\sigma_D} (\nabla I_D )^2,
\end{equation}
where $\sigma_D \equiv \sigma(I_D)$. The temporal dependence in \eqref{eq:thermal EP} is disregarded in the time-independent mesosocpic setup. Note that $\langle \Sigma \rangle $ is 
dimensionless, i.e. the mesoscopic counterpart of the Boltzmann factor $k_B$ is unity. 

Equipped with the definition of the disorder averaged cost function \eqref{eq: mean cost def}, we are now in a position  to prove the QMUR \eqref{qmur}. 
To that purpose, from the stochastic mesoscopic current density in (\ref{langevin}), we define the scalar quantity, 
\begin{equation}
 \label{eq: f particular}
 f \equiv \int d \mathbf{r} \,  \mathbf{j(\mathbf{r})} \cdot \mathbf{\widehat{n}}    
\end{equation}
where $\mathbf{\widehat{n}}$ is an arbitrary unit vector. Then, we define the inner product,
\begin{equation}
\label{eq:inner product}
\langle \mathbf{a} , \mathbf{b} \rangle_{\sigma_D} \equiv   \int d \mathbf{r} \,  \mathbf{a}\cdot \mathbf{b} \, \sigma^{-1} _D     
\end{equation}
which allows to write  $\langle f \rangle = \langle \mathbf{j}_D , \mathbf{\widehat{n}} \sigma_D \rangle_{\sigma_D}$ and $\langle f^2 \rangle_c  = \int d \mathbf{r} \, \sigma_D  = \langle \mathbf{\widehat{n}} \sigma_D , \mathbf{\widehat{n}} \sigma_D  \rangle_{\sigma_D}$ (see Appendix  \ref{app:cumulants of the radiative force}). 

The Cauchy-Schwarz relation associated to the inner product \eqref{eq:inner product} implies the inequality
\begin{equation}
\label{eq:CS onsager}
    \langle \sigma_D \mathbf{\widehat{n}} , \sigma_D \mathbf{\widehat{n}} \rangle_{\sigma_D}
    \langle \mathbf{j}_D , \mathbf{j}_D \rangle_{\sigma_D} \geq
    \langle \mathbf{j}_D , \mathbf{\widehat{n}} \sigma_D \rangle^2 _{\sigma_D},
\end{equation}
which together with $2 \langle \mathbf{j}_D,\mathbf{j}_D \rangle_{\sigma_D} = \langle  \Sigma \rangle  $ leads to the QMUR \eqref{qmur}. 

The linear dependence of $f$ upon  $\mathbf{j}$ may appear restrictive. Yet, it corresponds to a wealth of physically relevant mesoscopic quantities often considered, e.g. the force induced by coherent light fluctuations recently studied \cite{Soret19}. A generalised expression \eqref{eq:generalized QMUR}, for $f$ and the QMUR will be proposed in section~\ref{sec:Stochastic integral}.
We wish now to obtain an expression of the mesoscopic cost function $\Sigma$ at the stochastic level and not only as a disorder averaged quantity. It will allow to generalize the QMUR \eqref{qmur} and to include a corresponding large deviation bound. 


\section{Statistical field theory formulation }
\label{sec:Stochastic integral}

To implement this program, 
we first present a field theory description for the mesoscopic transport.

\subsection{ From Langevin equation to path probability  }

The Langevin equation \eqref{langevin} allows for a stochastic coarse grained approach of quantum mesoscopics, obtained by associating to each realization of the noise $\boldsymbol{\xi}(\mathbf{r})$, a path $\Gamma = \lbrace \mathbf{j}(\mathbf{r}) , I(\mathbf{r}) \rbrace $ with 
 a divergence free current $\nabla \cdot \mathbf{j} =0$  and appropriate boundary conditions \footnote{$\nabla \cdot \mathbf{j} $ does not necessarily vanish on the boundary. } (see appendix \ref{app:BC radiative transfer}). It would be tempting to identify the stochastic paths $\Gamma$ to the multiple scattering sequences ${\cal C}_N =
(\mathbf{r}_1, \mathbf{r}_2,..., \mathbf{r}_N)$ defined in section \ref{sec:Diffusion Equation}. This identification does not hold since ${\cal C}_N$ are microscopic scattering sequences obtained from a formal expansion of the Green's function of the Helmholtz equation (\ref{helmoltz}) for a given disorder configuration, while paths $\Gamma = \lbrace \mathbf{j}(\mathbf{r}) , I(\mathbf{r}) \rbrace $ are generated by the local stochastic noise (\ref{noise}), associated to quantum crossings, and correspond to  coarse grained trajectories.


It is useful to switch from the Langevin description to an equivalent statistical field theory. To that purpose, we employ the Martin-Siggia-Rose technique \cite{Martin73,tauber2014critical} to express the probability $P(\Gamma)$ of a path as
\begin{eqnarray}
    \label{eq:Field theory probability}
    P(\Gamma) \approx  \exp{\left[
    - \int d \mathbf{r} \,  \mathfrak{L}(\Gamma) \right]}
    \\ \nonumber 
    \mathfrak{L}(\Gamma)= \frac{(\mathbf{j}+D \nabla I)^2}{2\sigma(I) }
     .
\end{eqnarray}
The quadratic form of $\mathfrak{L}(\Gamma)$ results from  the (multiplicative) white noise $\boldsymbol{\xi}$ and from $\nabla \cdot \mathbf{j} =0 $ implicitly assumed in \eqref{eq:Field theory probability}.

The path probability \eqref{eq:Field theory probability} (as well as the Langevin equation \eqref{langevin}) is valid to leading order in $g_\mathcal{L}^{-1} \ll 1$ \footnote{The path probability \eqref{eq:Field theory probability}
is exact to leading order if $g_{\mathcal{L}}\gg 1 $. Otherwise, for $g_\mathcal{L} \sim 1$, subleading corrections to the quadratic $\mathfrak{L}(\Gamma)$ become relevant \cite{gardiner1985handbook,Lubensky_fieldTheory}.  }.  Moreover, 
in that limit, the path probability is dominated by a saddle point solution, so that   for any observable $O$
\begin{equation}
    \label{eq:saddle point}
    \langle O \rangle  = \int d\Gamma \, O(\mathbf{j},I) P(\Gamma) = O(\mathbf{j}_D, I_D). 
\end{equation}
Using \eqref{eq:Field theory probability} and \eqref{eq:saddle point}, it is now possible to define $\Sigma$ and show that $\langle \Sigma \rangle$ is given by \eqref{eq: mean cost def}.


\subsection{The Cost function}

We start by recalling some known results on the thermodynamic definition of the entropy production rate \cite{seifert2012stochastic,van2015ensemble}. Then, taking advantage of the analogy between non-equilibrium thermodynamics and quantum mesoscopics, we use the path probability to define the cost function.


Denoting by $P_{th} (\Gamma_{th})$ the path probability of a stochastic process, it is completely reversible if for any path $\Gamma_{th}$, the time-reversed path $\theta \Gamma_{th}$ is equally likely. With this intuition in mind, one can define the (dimensionless) entropy production variable
\begin{equation}
\label{eq:EPV}
\Sigma_{th} (\Gamma_{th}) =  \log \frac{P_{th}(\Gamma_{th})}{P_{th}(\theta \Gamma_{th})} .
\end{equation}
While $\Sigma_{th}$ can be negative, its average is non-negative,
\begin{equation}
    \label{eq:thermal EPV}
    \langle \Sigma_{th} \rangle_{th}  =   \int d\Gamma_{th} \, \left(
    P_{th}(\Gamma_{th}) - P_{th}(\theta \Gamma_{th})) \Sigma_{th}(\Gamma_{th}
    \right)  \geq 0 ,
\end{equation}
a result which stems from the non-negativity of the integrand, i.e. $(x-y)\log \frac{x}{y} > 0$ for $x,y> 0$. 

Analogously to \eqref{eq:EPV}, we define the cost function variable 
\begin{equation}
\label{eq:cost function variable}
\Sigma (\Gamma) \equiv \log \frac{P(\Gamma)}{P(\Theta \Gamma)},
\end{equation}
where $\Theta \Gamma \equiv \lbrace -\mathbf{j},I \rbrace$ is the reversed path. For the path $\Gamma$ to exist with non-vanishing probability, it needs to correspond to some realization of the noise $\boldsymbol{\xi}$ satisfying \eqref{langevin}, to hold the boundary conditions (see appendix \ref{app:BC radiative transfer})  and maintain $\nabla \cdot \mathbf{j}=0$. If $\Gamma$ exists, so does $\Theta \Gamma$: $\nabla \cdot (-\mathbf{j})$ vanishes, $\boldsymbol{\xi} \rightarrow \boldsymbol{\xi} - 2\mathbf{j} $ satisfies \eqref{langevin} and the boundary conditions apply (see appendix \ref{app:BC radiative transfer}).

From the path probability \eqref{eq:Field theory probability} and the cost function variable \eqref{eq:cost function variable} we find $\Sigma = -\int d \mathbf{r}  \, \frac{2 \mathbf{j} \cdot D \nabla I}{\sigma(I)} $. Using the saddle point approximation \eqref{eq:saddle point},  
\begin{equation}
\label{eq: cost after averaging}
\langle \Sigma \rangle  = - \int d \mathbf{r} \frac{2 \mathbf{j}_D \cdot D \nabla I_D}{\sigma_D} =   \int d \mathbf{r} \frac{2  D^2 ( \nabla I_D)^2}{\sigma_D}.
\end{equation}
Therefore, the disorder averaged cost function variable corresponds to the cost function defined in \eqref{eq: mean cost def}. 

While we have discussed so far the analogs between thermal and quantum mesoscopic systems, it is important to note that the underlying physics is quite different. For example, entropy production is notoriously hard to measure in thermal systems \cite{martinez2019inferring}. Here, we want to argue that the cost function is accessible experimentally.  To do so, note that $\langle \Sigma \rangle $ depends on $I_D$ alone (through $D,\sigma$). Since $I_D$ is a solution of a Laplace equation, it is completely determined by the boundary conditions and therefore so does $\langle \Sigma \rangle $. Let us express the relation of the cost function \eqref{eq: cost after averaging} to the boundary conditions;    



\begin{equation}
\begin{array}{ll}
\langle \Sigma\rangle &= \frac{D^2}{c_0}\int_V d \mathbf{r}\, \frac{[\nabla I_D(\mathbf{r})|^2}{I_D(\mathbf{r})^2}\\
\\
&= \frac{D^2}{c_0}\int_V d \mathbf{r}\, \nabla\left(\frac{-1}{I_D(\mathbf{r})}\right)\cdot\nabla I_D(\mathbf{r})
\end{array}
\end{equation}
with $c_0 = \frac{2\pi l v^2}{3k^2}$. Further integration by parts yields
\begin{equation}
\begin{array}{ll}
\langle \Sigma\rangle = \frac{D^2}{c_0}\oint_S \frac{-1}{I_D(\mathbf{r})}\nabla I_D(\mathbf{r})\cdot d\mathbf{S} - \frac{D^2}{c_0} \int_V d \mathbf{r}\, \frac{-1}{I_D(\mathbf{r})}\Delta I_D(\mathbf{r})
\end{array}
\label{sigma-aux}
\end{equation}
where $d\mathbf{S} = dS \, \mathbf{\widehat{n}}(\mathbf{r})$ with $\mathbf{\widehat{n}}(\mathbf{r})$ the normal vector to the infinitesimal surface $dS$ located at the point $\mathbf{r}$ on the boundary. The second term of the right hand side of \eqref{sigma-aux} vanishes since $-\Delta I_D = 0$. Let us rescale the surface integral in the right hand term of \eqref{sigma-aux} by the characteristic length of the system $\mathcal{L}$, i.e.   $\mathbf{\tilde{r}} = \mathbf{r}/\mathcal{L}$ and $\mathbf{\tilde{S}} = d\mathbf{S}/\mathcal{L}^2$.  We find 
\begin{equation}
    \begin{array}{ll}
\langle \Sigma\rangle &= - \frac{k^2\mathcal{L} l}{6\pi}\oint_{\partial \tilde{V}} \frac{1}{I_D(\mathbf{\tilde{r}})}\nabla I_D(\mathbf{\tilde{r}})\cdot d\mathbf{\tilde{S}}
= g_{\mathcal{L}} \mathcal{B} ,
\\ \label{sigma}
\mathcal{B} &= -\frac{1}{2}\oint_{\partial \tilde{V}} \frac{1}{I_D(\mathbf{\tilde{r}})}\nabla I_D(\mathbf{\tilde{r}})\cdot d\mathbf{\tilde{S}}
    \end{array}
\end{equation}
Here $\mathcal{B}$ depends only on the boundary conditions. 
Note that $\Theta \Gamma$ is not a time reversed path. Indeed, in the case of a uniformly illuminated and symmetric sample, $I_D$ is uniform and therefore $\langle\Sigma\rangle=0$. However, quantum crossings still occur and TRS is still broken.



\subsection{Generalized expression of the QMUR}

Having obtained expression \eqref{eq: cost after averaging} for the cost function before disorder averaging, by means of the path probability \eqref{eq:Field theory probability}, we are now in a position to generalize the
QMUR in \eqref{qmur} by relaxing the linear dependence of $f$ defined in \eqref{eq: f particular}. 
To that purpose, we consider the generalized expression 
\begin{equation}
f(\Gamma) =
 \int d \mathbf{r} \, 
 z( \Gamma),
\end{equation}
 for $f$ with $z$ an arbitrary function. We wish to explore how the fluctuations of $f$ are bounded. To that end, we define the cumulant generating function of $f$,
\begin{equation}
    \mu(\lambda) = \log \langle e^{ \lambda f} \rangle \approx \int d\Gamma e^{\lambda f(\Gamma) - \int d \mathbf{r} \mathfrak{L}(\Gamma)}. 
    \label{cgf-eq}
\end{equation}
Next, we derive in the spirit of \cite{Sasa18}, the QMUR and its generalization to the cumulant generating function. To do so, we consider another path probability defined with the tilted Lagrangian $\mathfrak{L}_\mathbf{Y} = (\mathbf{j}+D\nabla I - \mathbf{Y})^2 / 2\sigma $, where $\mathbf{Y}$ is a divergence free field. The tilted path probability corresponds to the Langevin dynamics $\mathbf{j} = -D\nabla I + \mathbf{Y} + \boldsymbol{\xi}$, with the same noise $\boldsymbol{\xi}$ defined in \eqref{langevin}. The tilted process disorder average is denoted by $\langle \cdot \rangle_{\mathbf{Y}} $, such that  $\langle \cdot \rangle_{\mathbf{0}} =  \langle \cdot \rangle $. The usefulness of the tilted dynamics comes from the fact that under the tilted disorder average, the intensity remains unchanged, i.e. $\langle I \rangle_\mathbf{Y}  = I_D  $ for any divergence free field $\mathbf{Y}$, but the average current gets a tilt, i.e. $\langle j \rangle_\mathbf{Y} = \mathbf{j}_D + \mathbf{Y}   $. 
This tilting dynamics has been used to create a mapping to equilibrium and to generate the time-reversed dynamics in the thermal case \cite{Sasa18,Shpielberg_Pal}. 
Here we use it to optimize a bound on $\mu(\lambda)$.   
Using the identity 
\begin{equation}
    \label{eq:tilted Lagrangian relation}
    \mathfrak{L}= \mathfrak{L}_{\mathbf{Y}}+ |\mathbf{Y}|^2 / 2\sigma +  \frac{\mathbf{Y}}{\sigma} \cdot (\mathbf{j}+D \nabla I  - \mathbf{Y} ) ,
\end{equation}
allows to rewrite the cumulant generating function $\mu(\lambda)$ as
\begin{equation}
    \mu(\lambda ) = \log \langle e^{\lambda f - \int d \mathbf{r} \, 
    |\mathbf{Y}|^2 / 2\sigma + \frac{\mathbf{Y}}{\sigma} \cdot (\mathbf{j}+D \nabla I  - \mathbf{Y} )} \rangle_\mathbf{Y} .
\end{equation}
The Jensen inequality, $ \langle e^{ O }  \rangle_\mathbf{Y} \geq e^{\langle O \rangle_\mathbf{Y} }  $, 
 then implies 
\begin{equation}
\label{eq:after Jensen}
    \mu(\lambda) \geq \lambda \langle  f \rangle_\mathbf{Y} - \int d \mathbf{r} \, \langle \frac{|\mathbf{Y}|^2}{2\sigma } \rangle_{\mathbf{Y}} ,
    \end{equation}
noting that the term $  \frac{\mathbf{Y}}{\sigma} \cdot    (\mathbf{j}+D \nabla I  - \mathbf{Y} ) $ vanishes under the tilted disorder averaging. Choosing $\mathbf{Y} = \alpha \mathbf{j}_D$ and noting that   the tilting field leaves the disorder averaged intensity $I_D$ unchanged, 
we find 
\begin{eqnarray}
\label{eq: EPR identifications}
    \langle \frac{2\mathbf{j}^2 _D}{\sigma } \rangle_{\alpha \mathbf{j}_D} &=& \langle \Sigma  \rangle 
    \\ \nonumber 
    \langle f \rangle_{\alpha \mathbf{j}_D} &=& \int d \mathbf{r} \, z((1+\alpha)\mathbf{j}_D, I_D). 
\end{eqnarray}
From \eqref{eq:after Jensen} and \eqref{eq: EPR identifications}, we recover the cumulant generating function bound 
\begin{equation}
    \label{eq:CGF bound}
    \mu(\lambda) \geq  \lambda \langle f \rangle_{\alpha \mathbf{j}_D}  - \frac{1}{4}\alpha^2 \langle \Sigma \rangle.
\end{equation}
This inequality is valid for any $\alpha$ and any choice of $f$. To recover the generalized QMUR, we assume $\alpha \ll 1$ and develop the right hand side of \eqref{eq:CGF bound} to second order in $\alpha$. The quadratic expression can be optimized by $\alpha = 2\lambda  \langle \partial_{\mathbf{j}} f \rangle / \langle \Sigma\rangle  $, where 
\begin{equation}
    \langle \partial_{\mathbf{j}} f \rangle = \int d \mathbf{r} \, \mathbf{j}_D  \frac{\delta }{\delta \mathbf{j}_D} z(\mathbf{j}_D,I_D).
\end{equation}
Then, the inequality to second order in $\lambda$ implies the generalized QMUR 
\begin{equation}
    \label{eq:generalized QMUR}
    \mu''(0) = \langle f^2 \rangle_c  \geq \frac{2 \langle \partial_{\mathbf{j}} f \rangle^2  }{\langle \Sigma \rangle }.
\end{equation}
Using the optimal $\alpha$, we recover a large deviation bound for the fluctuations of $f$. Namely, using $\alpha = 2 \lambda \langle \partial_{\mathbf{j}} f \rangle / \langle \Sigma \rangle $ in \eqref{eq:CGF bound}. For example, the linear choice $z = \mathbf{j} \cdot \widehat{n}$ leads to
\begin{equation}
\label{eq:CGF bound f}
    \mu(\lambda) \geq \lambda \langle f \rangle  + \frac{\langle f \rangle^2 }{\langle \Sigma\rangle}.
\end{equation}

In this section, we have introduced the definition of the stochastic cost function $\Sigma$. We have also proved the QMUR for general fluctuating quantities $f$ in \eqref{eq:generalized QMUR} and derived a large deviation bound \eqref{eq:CGF bound} and \eqref{eq:CGF bound f}. Note that $\langle \Sigma \rangle $ arises naturally in the optimization of the QMUR. This comes from the coarse grained level of the Langevin equation, leading to a quadratic Lagrangian $\mathfrak{L}$. We note that for a microscopic theory with non-quadratic Lagrangian, e.g. for a master equation, the optimal bound is much more cumbersome and currently lacks physical interpretation \cite{shiraishi2021optimal}. 
We now apply the QMUR to some physically relevant examples.


\section{Examples}
\label{sec:Examples}

Calculating explicitly the disorder averaged intensity for an arbitrary sample usually requires a numerical approach. Moreover, careful preparation of experimental samples also requires a simple setup. The slab geometry, represented on Fig.~\ref{fig-dis-rev}, has the double advantage of being both experimentally accessible and analytically solvable. For the setup of Fig.~\ref{fig-dis-rev}, the disorder averaged intensity is linear, 
\begin{equation}
    I_D(x) = \frac{I_0}{4\pi}\frac{-5+5e^{-L/l}}{4l/3+L}x + \frac{I_0}{4\pi}\frac{5L+\frac{10l}{3}(1+e^{-L/l})}{4l/3+L} \, ,
\end{equation}

see appendix \ref{app:slab geometry}.

We focus on two important mesoscopic fluctuating quantities; The transmission coefficient and fluctuation induced radiative forces. We check that these forces satisfy the QMUR. Then, the fluctuation induced radiative forces are shown numerically to satisfy the large deviation bound \eqref{eq:CGF bound f}.

\subsection{QMUR for the transmission coefficient}

First, we present a direct check of the generalised QMUR \eqref{eq:generalized QMUR} for the transmission coefficient $T(\theta)$ -- the ratio between the light intensity transmitted in the direction $\mathbf{\widehat{s}}$ (see Fig.~\ref{fig-dis-rev}) and the incoming intensity. The transmission coefficient and its fluctuations -- which give rise to speckle patterns -- have been extensively studied and measured \cite{Goodman,Akkermans,Boer92,Stephen87}. Deciphering the information encoded in fluctuations of the transmission coefficient is still an active field of research, with a broad range of applications such as imaging of biological tissues and turbid media, sensing and information transmission in random media \cite{Fayard18,Sarma19}. Remarkable progress has also been made recently in the ability to control the transmission of light in random media, with the emergence of wave front shaping techniques \cite{Mosk07,Rotter17}. In this context, providing a general bound for the fluctuations of $T(\theta)$ using the QMUR is of particular interest.

Let $I(\mathbf{\widehat{s},r})$ be the fraction of light intensity propagating in the direction $\mathbf{\widehat{s}}$. The transmission coefficient is then defined as $T(\theta) = \widehat{s}_x I(\mathbf{\widehat{s}},L)/I_0$, where $\theta\in[-\pi/2,\pi/2]$ is the angle between $\mathbf{\widehat{s}}$ and the $x$ axis. In the literature, $I(\mathbf{\widehat{s},r})$ is called the specific intensity \cite{Ishimaru,Akkermans}. Its angular average gives the total intensity, $I (\mathbf{r}) = \overline{ I (\mathbf{\widehat{s},r})} $.
The specific intensity satisfies the radiative transfer equation. For details on how to obtain this equation, we refer the reader to the section A5.2 in \cite{Akkermans}. In the absence of light source inside the medium, it is given by 
\begin{equation}
    I(\mathbf{\widehat{s}},x) = I(x) + \frac{l}{D}\mathbf{\widehat{s}}\cdot\mathbf{j}(x).
\end{equation}
We can therefore write $T(\theta)$ in the form 
\begin{equation}
T(\theta) = \frac{1}{S}\int_V d\mathbf{r}\, \delta(x-L) \widehat{s}_x \frac{I(x) + \frac{l}{D}\mathbf{\widehat{s}}\cdot\mathbf{j}(x)}{I_0} 
\end{equation}
where 
$z(I,\mathbf{j}) = \delta(x-L) \widehat{s}_x (I(x) +\frac{l}{D}\mathbf{\widehat{s}}\cdot\mathbf{j}(x))/I_0 $, and use \eqref{eq:generalized QMUR} to obtain
\begin{equation}
    \langle T^2(\theta)\rangle_c \langle\Sigma\rangle\geq 2  \langle\partial_{\mathbf{j}} T(\theta)\rangle^2,
    \label{qmur-T}
\end{equation}
where the lower bound of the QMUR is given by
\begin{equation}
\begin{array}{ll}
    \langle\partial_{\mathbf{j}} T(\theta)\rangle &= \langle \frac{1}{S}\int\limits_V d\mathbf{r}\,\delta(x-L) \frac{\widehat{s}_x l}{DI_0}\mathbf{\widehat{s}}\cdot\mathbf{j}\rangle \\
    \\
    &= \frac{\widehat{s}_x l}{DI_0}\frac{1}{S}\int\limits_S d\mathbf{S}\, \mathbf{\widehat{s}}\cdot\mathbf{j}_D.
\end{array}
\label{lower-bound}
\end{equation}
Using Fick's law $\mathbf{j}_D = -D\nabla I_D$ and the solution to \eqref{id-slab} for $I_D$ in a slab geometry, we obtain the expression of the lower bound, $\langle\partial_{\mathbf{j}} T(\theta)\rangle = \widehat{s}_x^2\frac{15(1-e^{-L/l})}{8\pi(u+2)} \simeq \widehat{s}_x^2\frac{15}{8\pi(u+2)}$ where $u=3L/2l$ and $\widehat{s}_x = \cos(\theta)$. The lower bound reaches its maximum for $\theta=0$. In the slab geometry, the correlation function of the transmission coefficient is given by (see section 12.4 in \cite{Akkermans}) 

\begin{equation}
 \langle T^2(\theta)\rangle_c = \frac{4}{3g_\mathcal{L}} \left( \frac{15 \widehat{s}_x}{8 \pi u} \right)^2 \left(\widehat{s}_x + 2/3 \right)^2.
\end{equation}
Reinjecting this expression, together with the lower bound \eqref{lower-bound} and $\langle \Sigma\rangle =g_\mathcal{L}\frac{u^2}{(1 + u )}$, in the QMUR \eqref{qmur-T}, and rearranging the terms to separate those depending on $u$ and $s_x$, we find

\begin{equation}
     \frac{2}{3} \left( \frac{ \cos \theta + 2/3}{\cos \theta}  \right)^2 \geq \frac{1 + u }{( 2 + u )^2}\, . 
     \label{sat-cond}
\end{equation} 
For $u>0$ and $\theta\in [-\pi/2,\pi/2]$, we have
\begin{equation}
\frac{1 + u }{( 2 + u )^2}  \leq 1/4    \quad \textrm{and} \quad \frac{2}{3} \left( \frac{ \cos \theta + 2/3}{\cos \theta}  \right)^2 \geq 50/27.
\end{equation}
 Hence \eqref{sat-cond} is always satisfied, and the QMUR \eqref{qmur-T} is indeed justified. 

\subsection{QMUR for radiative forces}

We now briefly discuss the recently studied radiative force induced by mesoscopic coherent fluctuations of light \cite{Soret19}. The radiative force exerted on a suspended membrane, of surface $\delta S$, immersed in the medium, see Fig.~\ref{fig-dis-rev}.b, is given by $\delta\mathbf{f}=\mathbf{\widehat{n}} \, v^{-2}\int_{\delta S}\mathbf{j}\cdot\mathbf{\widehat{n}}$ where $\mathbf{\widehat{n}}$ is a unit vector normal to $\delta S$ and $v$ is the group velocity. 
As a result of coherent effects described by quantum crossings, this force displays fluctuations, which typically scale like $\langle \delta f^2\rangle_c/\langle \delta f\rangle^2\sim 1/g_\mathcal{L}$ \cite{Soret19}. Since $g_\mathcal{L}$ is an easily tunable parameter, one can choose a setup where the fluctuations are measurable, and significantly enhanced compared to other forces exerted on the membrane, such as Van der Waals forces \cite{Soret19}. 
The spatially averaged force, $f_{av}= v^{-2}\int_V\mathbf{j}\cdot\mathbf{\widehat{n}}$, satisfies the QMUR \eqref{qmur}. Indeed using $\langle f_{av}^2\rangle_c= v^{-4}\int_V \sigma_D (\mathbf{r})$ and the expressions for $I_D$ and $\mathbf{j}_D$ in a slab geometry, given earlier, we obtain 
\begin{equation}
\begin{array}{ll}
\frac{\langle f_{av}^2\rangle_c\langle\Sigma\rangle }{2\langle f_{av} \rangle^2} &= \frac{(u+1)^3 - 1}{ 3u(u+1)}  \geq 1.
\end{array}
\label{ratio}
\end{equation}
where again $u=3L/2l$.
Equality in \eqref{ratio} is attained only in the nonphysical $u=0$ value; experimentally, it is reasonable to achieve $u\sim 10$, for which the ratio (\ref{ratio}) is $\sim 4$. Indeed we find that the QMUR \eqref{ratio} is a good bound on the fluctuation induced force inside the slab.  

\subsection{Cumulant generating function for radiative forces}

\begin{figure}
    \centering
    \includegraphics[scale=0.5]{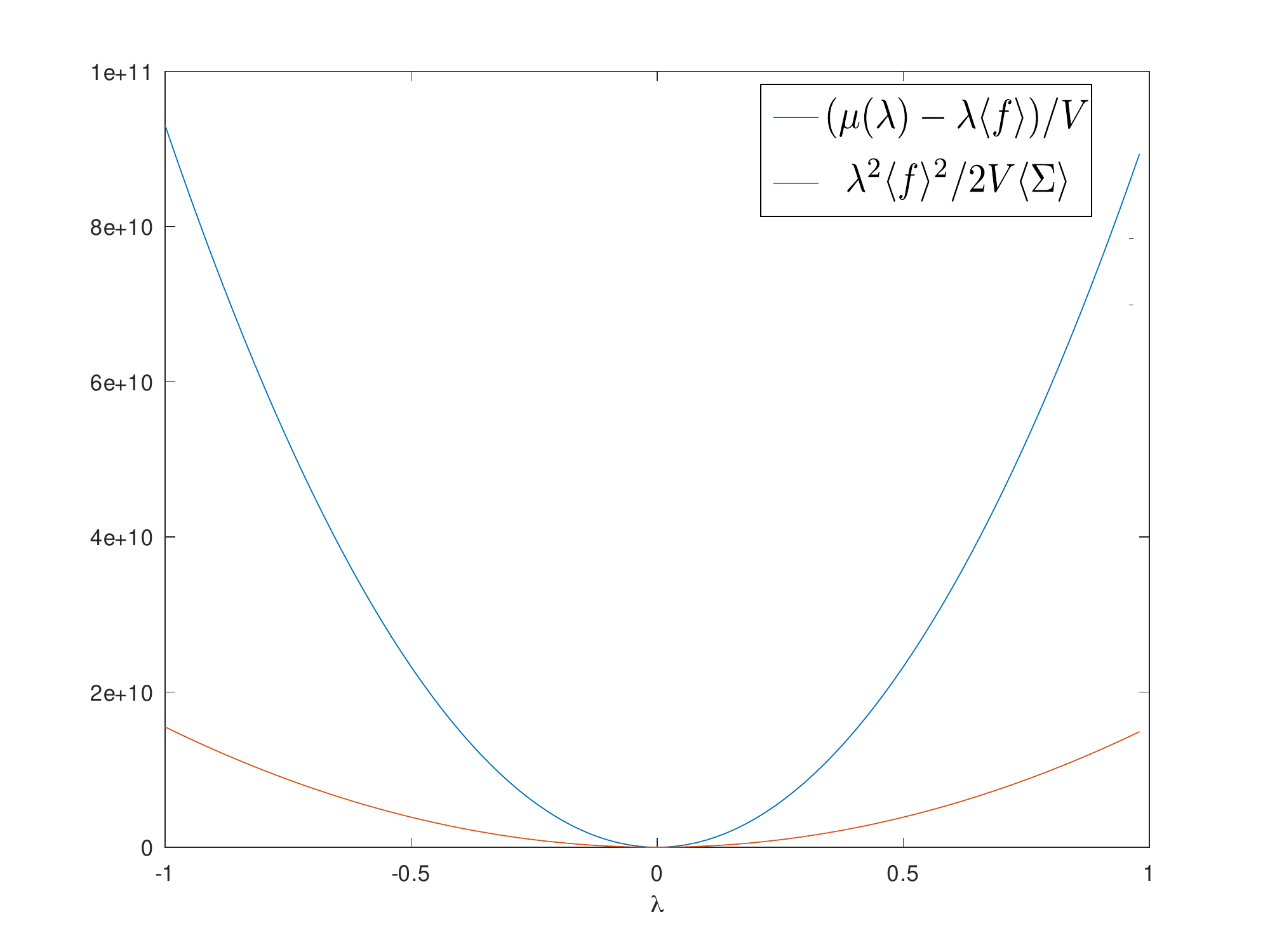}
    \caption{ Numerical verification of the lower bound to the cumulant generating function \eqref{eq:CGF bound f} of the fluctuation induced coherent force for the slab geometry. The figure presents  $\mu(\lambda)- \lambda \langle f_{av} \rangle $ (blue) and its lower bound  $\lambda^2 \langle f_{av} \rangle ^2 / \langle \Sigma \rangle  $ (red), both divided by the volume. Here we consider $L/l_0=5$.  The bound is tightest in the slab geometry when $L/l_0$ is as small as physically possible.} 
    \label{cgf}
\end{figure}

Finally, we check numerically the inequality (\ref{eq:generalized QMUR}) for the radiative forces,  $f_{av} = v^{-2} \int d \mathbf{r} \, \mathbf{j}(\mathbf{r}) \cdot \mathbf{\widehat{n}} $. To compute (\ref{cgf-eq}), we use the fact that, since the noise is weak, the integrand on the r.h.s. of (\ref{cgf-eq}) dominated by the saddle point solution \eqref{eq: Hamilton equations}. We obtain the saddle point solution \eqref{eq: Hamilton equations} numerically, and check that the cumulant generating function satisfies (\ref{eq:generalized QMUR}), see Fig.~\ref{cgf}.

\section{Cram\'er-Rao bound and Fisher Information}

The purpose of this section is to rederive the QMUR using the Cram\'er-Rao bound, identifying the cost function $\langle \Sigma \rangle$ as the Fisher information. 

The Fisher information is a way of measuring the amount of information that an observable random variable $\Gamma$ carries about an unknown parameter $\theta$ upon which the probability of $\Gamma$ depends.   The Cram\'er-Rao bound is given for any function $\zeta(\Gamma)$, 
\begin{equation}
\label{eq:Cramer Rao general bound}
    \frac{\mathrm{Var}_\theta \left[ \zeta(\Gamma) \right]}{(\partial_\theta \langle \zeta(\Gamma) \rangle_\theta )^2} \geq 1/ \mathcal{I}(\theta). 
\end{equation}
We can prove the Cram\'er-Rao bound using the Cauchy-Schwarz inequality, and will do so later on, in \ref{prvoing CR bound}.

To apply the Cram\'er-Rao bound to the mesoscopic case, let us consider the tilted diffusion  $D_\theta = D e^\theta$. Furthermore, we define the probability $P_\theta (\Gamma)  $ by replacing $D \rightarrow D_\theta$. 
This implies replacing $\mathfrak{L} \rightarrow \mathfrak{L}_\theta = (j- e^\theta J)^2 / 2\sigma $. We define the Fisher information $\mathcal{I}(\theta) = \langle (\partial_\theta \log P_\theta )^2 \rangle_\theta$. 

One can then show that $\mathcal{I}(0) = \langle \Sigma \rangle$. Then, it is simple enough to show that setting $\zeta = \int d\mathbf{r} \, \mathbf{j} \cdot \mathbf{\widehat{n}} $ leads to the QMUR in \eqref{qmur}. 
What we have gained here is an interpretation of the cost function $\langle \Sigma \rangle$ as the Fisher information of changing the diffusion coefficient. 

\subsection{Proving the general Cram\'er-Rao bound \label{prvoing CR bound}}

Let us define for the function $\zeta(\Gamma)$, $\psi(\theta) \equiv \langle \zeta(\Gamma) \rangle_\theta$. Furthermore, we define the inner product $\langle a , b \rangle_\theta = \int d\Gamma a(\Gamma) b(\Gamma) P_\theta (\Gamma)  $.   We notice that 
\begin{equation}
    \langle (\zeta(\Gamma)-\psi(\theta)) , \partial_\theta \log P_\theta \rangle_\theta = \partial_\theta \psi(\theta). 
\end{equation}
Then, applying the Cauchy-Schwarz inequality, we find 
\begin{equation}
|\langle \zeta(\Gamma)-\psi(\theta),\zeta(\Gamma)-\psi(\theta) \rangle_\theta |^2  |\langle \partial_\theta \log P_\theta,\partial_\theta \log P_\theta \rangle_\theta |^2      \geq  (\partial_\theta \psi(\theta) )^2 .
\end{equation}
Identifying 
\begin{eqnarray}
\langle \Theta(\Gamma)-\psi(\theta),\zeta(\Gamma)-\psi(\theta) \rangle_\theta &=& \mathrm{Var}_\theta \left[ \zeta(\Gamma)\right]  \\ \nonumber 
\langle  \partial_\theta \log P_\theta,\partial_\theta \log P_\theta  \rangle_\theta  &=& \mathcal{I}(\theta)
\end{eqnarray}
we recover \eqref{eq:Cramer Rao general bound}.

\section{Discussion/Conclusion}

The recently discovered TUR reveal a universal bound on precision of thermal machines given by the entropy production. The TUR demonstrates that there are limits to what could be simultaneously achieved in a stochastic system. 

Not all stochastic system are thermal. Therefore, it stands to reason that the TUR could be generalized to athermal systems. A major difficulty to achieving this goal comes from the fact that while there have been attempts to generalize the notion of entropy to athermal systems \cite{Baule_Review,Levine_Data}, there are no such generalizations to entropy production. In this work, we proved the QMUR -- a generalization of the TUR to zero temperature quantum mesoscopic physics. Here fluctuating quantities, e.g. fluctuation induced forces and fluctuating transmission coefficients, arise from coherent terms, i.e. wave interference. The cost function, generalizing the entropy production rate, has been defined as the log of the ratio between the path probability $P(\lbrace \mathbf{j}(\mathbf{r}),I(\mathbf{r}) \rbrace)$  and the reversed path probability $P(\lbrace -\mathbf{j}(\mathbf{r}),I(\mathbf{r}) \rbrace)$. 

Two comments are now in order. First, note that the QMUR was proved in three unrelated ways. Both in the field theory description as well through the Cram\'er-Rao bound, the cost function emerges naturally from the optimization of the bound. 
Despite the rich literature in the field, the cost function was never addressed. Nevertheless, the emergence of the cost function as a bound on coherent fluctuations implies it is an important mesoscopic quantity. Second, there are setups for which the current $\mathbf{j}_D$ vanishes, e.g. if the slab of Fig.~\ref{fig-dis-rev} were illuminated on both sides with the same intensity $I_0$. In this case, the average cost function (\ref{eq: cost after averaging}) vanishes. However, quantum crossings are still present, and hence, as argued in section 3, TRS is broken. Therefore, the cost function $\Sigma$ does not measure the breaking of time-reversal, and $\Theta\Gamma$ is not the time-reversed path.

The cost function $\Sigma$ serves as a bound on the coherent contributions --  the analog for precision. Furthermore, the QMUR was extended to a large deviation bound, again in terms of the cost function \eqref{eq:CGF bound f}. We have demonstrated the validity of the QMUR for two important measurable quantities: the fluctuating transmission coefficient and the coherent fluctuating induced force. We stress that analytic solution exists for simple setups, e.g. the slab geometry, calculating the fluctuating properties for an arbitrary setup is a non-trivial task. Hence arises one useful aspect of the QMUR, estimation of the coherent fluctuations in terms of the incoherent intensity $I_D$ alone.

Beyond these fundamental implications, our findings have a threefold interest. First, the QMUR \eqref{qmur} and \eqref{eq:generalized QMUR} provide a way to monitor coherent light fluctuations using the cost function $ \Sigma $ and its dependence upon boundary conditions through $\mathcal{B}$, and not only the dimensionless  conductance $g_{\mathcal{L}}$. Increasing coherence, especially through the boundary geometry, is of practical importance as current fluctuations are used as probes in biology and soft matter physics \cite{Mosk12} \cite{Cox}. 
Secondly, importing methods from statistical mechanics to mesoscopic physics, such as uncertainty relations \cite{Barato15,Gingrich2016,HorowitzReview} and lower bounds for the fluctuations, may prove helpful for imaging and wave control in complex media \cite{Gigan14,Fayard15,Fayard18}. Finally, in thermal systems it is often hard to measure entropy production and to determine the conditions for a tight bound of TUR.
Conversely, the significant progress made in recent years in the ability to control the light flow in random media \cite{Mosk07,Rotter17,Vellekoop10}, paves the way for experimental verification of QMUR and measurements of the mesoscopic cost function.

We also wish to stress that the present Langevin description applies beyond the case of scalar coherent light propagation so as to include e.g  polarization effects, anisotropic scattering and electronic quantum transport. But, extending the applicability of QMUR close to a Anderson localisation transition (i.e. for $g_\mathcal{L} \sim 1 $) where the Langevin approach is expected to break down appears more challenging. Yet, noting the unexpected  connection between the cost function and Fisher information \cite{Hasegawa19_TUR_Fisher,Ito_Dechant_TUR_Geometry,Pal2020} is a possible path to explore to study  QMUR for $g_\mathcal{L} \sim 1$. Finally, in this work we restricted ourselves to $1/g_\mathcal{L}$ corrections. Investigating whether a cost function and a resulting QMUR exist if we include $1/g_\mathcal{L}^2$ corrections is an open question.



\begin{acknowledgements}
M. Goldstein and N. Fayard are acknowledged for fruitful discussions.
\end{acknowledgements}

\renewcommand{\thesection}{\Roman{section}} 
\setcounter{section}{0}

\section{Radiative Transfer Equation and Boundary Conditions}
\label{app:BC radiative transfer}

In this section, we discuss the boundary conditions for the diffusive intensity $I_D$. The exact boundary conditions (\ref{prob-1}) for multiply scattered light intensity are not trivial, and we refer to the section A5.2 in \cite{Akkermans} for a detailed derivation. Moreover, in this exact description, the light intensity satisfies a diffusion equation with a source term, unlike the convention used in the main text, where we assumed $-D\Delta I_D = 0$. The purpose of this section is to obtain an alternative set of boundary conditions (\ref{prob-2}), which, associated with the source free diffusion equation, give a good approximation for the intensity $I_D$,  simplifying the derivation of the QMUR. 

\medskip

The idea behind the boundary conditions for the diffusive light intensity is to formalize that, since diffusive processes happen inside the disordered medium, there can be no incoming diffusive intensity at the interface. For a random medium, illuminated by an external light source of intensity $I_0$, propagating in the direction $\mathbf{\widehat{k}}$, the diffusive intensity is the solution of the following problem, 
\begin{equation}
\begin{array}{ll}
\Delta I_D(\mathbf{r}) = -\frac{v}{Dl}I_0(\mathbf{r})\\
\\
 I_D(\mathbf{r}) + \frac{2l}{3}\mathbf{\widehat{n}}\cdot\nabla I_D(\mathbf{r}) = 0 \mbox{ for any $\mathbf{r}$ at the interface}
\end{array}
\label{prob-1}
\end{equation}
where $\mathbf{\widehat{n}}$ is the normal unit vector at the point $\mathbf{r}$ on the surface. $I_0(\mathbf{r}) $ is the ballistic component of the intensity, corresponding to the fraction of the incoming radiation which propagates without any collisions on the scatterers; it decays exponentially with the distance to the surface, $I_0(\mathbf{r}) \propto  e^{-\mathbf{r}\cdot\mathbf{\widehat{k}}/l}$. 

\begin{figure}
    \centering
    \includegraphics[scale=0.3]{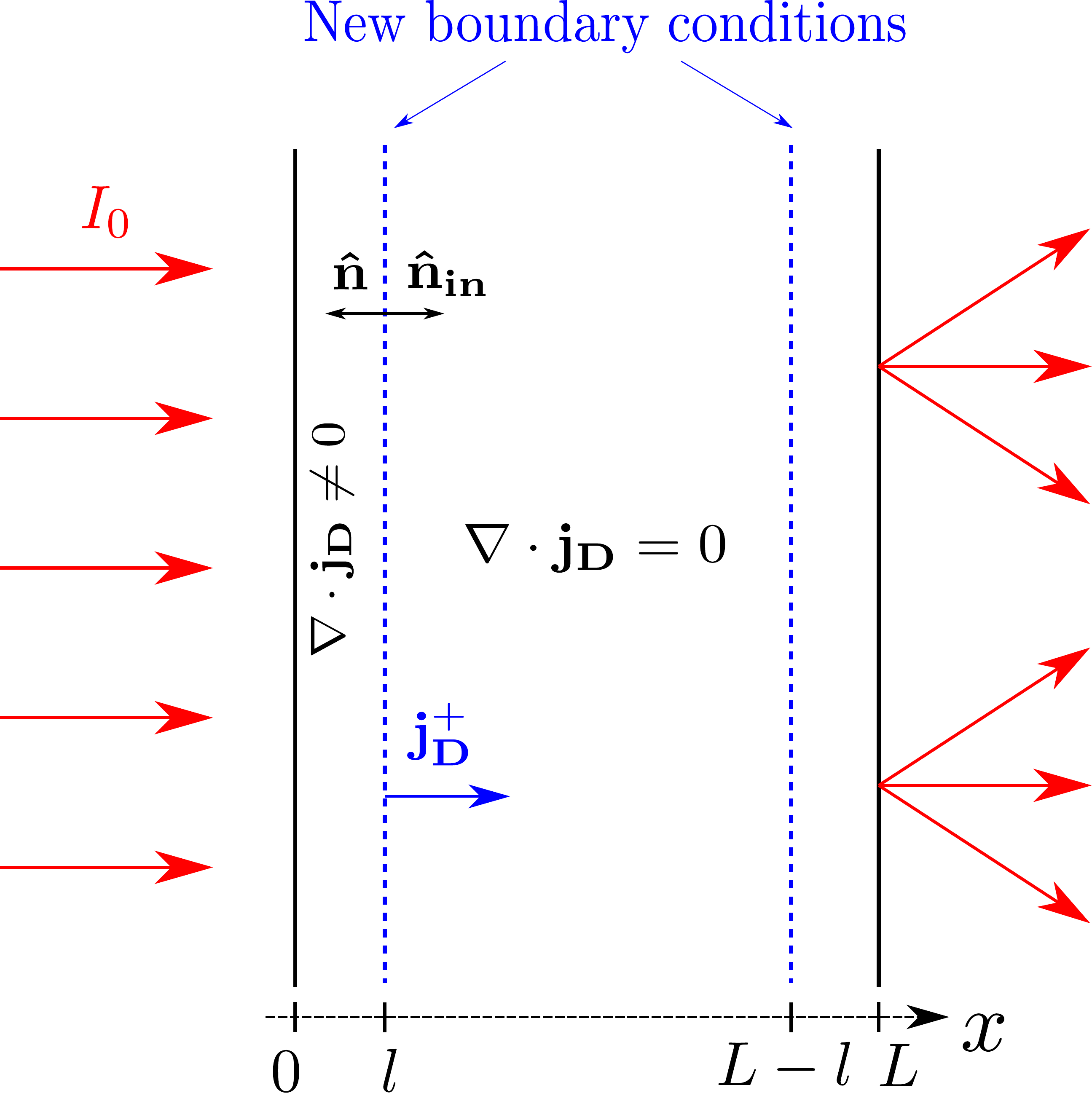}
    \caption{A slab of scattering medium is illuminated by a plane wave. The diffusive current obeys a continuity equation $\nabla\cdot\mathbf{j}_D=vI_0(\mathbf{r})/l\simeq 0$ for $\mathbf{r}$ at a distance $>l$ from the boundary. We solve for $I_D,\mathbf{j}_D$ in the bulk, assuming the current to be divergence free, and shifting the boundary conditions to the fictive boundary defined as the surface at a distance $l$ from the boundary (blue doted lines). }
    \label{fig-bc}
\end{figure}

We wish to reformulate the boundary conditions in order to have a source-less diffusion equation for $I_D$ -- or equivalently $\nabla\cdot\mathbf{j}_D = 0$ -- which is more convenient for the derivation of the QMUR in the main text. We begin by noticing that $\Delta I_D(\mathbf{r}) \simeq 0$ for $\mathbf{r}$ at a distance $>l$ from the surface. The idea is to neglect the layer of width $l$ at the boundary, and to solve for $I_D,\mathbf{j}_D$ in the bulk, where we can assume $\Delta I_D = 0$, and impose as boundary conditions the solutions of the exact problem (\ref{prob-1}) at the distance $l$ from the boundary, see Fig.~\ref{fig-bc}.

To avoid confusion, we note $I'_D,\mathbf{j'}_D$ the approximated solutions in the bulk, such that $\nabla\cdot\mathbf{j'}_D =0$.

We obtain the boundary conditions for $I'_D,\mathbf{j'}_D$ by calculating the incoming current $j_{D}^+=\mathbf{j_{D}^+}\cdot\mathbf{\widehat{n}_{in}}$ of the real problem (\ref{prob-1}) at the distance $l$ from the boundary. By definition, $j_D^+ = v\langle \mathbf{\widehat{s}}\cdot\mathbf{\widehat{n}_{in}} I_D(\mathbf{\widehat{s},r)}\rangle_\mathbf{\widehat{s}^+}$ where the average is taken over the half space $\mathbf{\widehat{s}}\cdot\mathbf{\widehat{n}_{in}}\geq 0$. On the other hand, $j_D^+$ is related to $I_D$ by means of the radiative transfer equation \cite{Akkermans},
\begin{equation}
\begin{array}{ll}
    I_D(\mathbf{\widehat{s},r}) &= I_D(\mathbf{r})-l\mathbf{\widehat{s}}\cdot\nabla I_D(\mathbf{r})\\
    \\
    \Rightarrow j_D^+(\mathbf{r})  &= \frac{v}{2}I_D(\mathbf{r}) - \frac{vl}{3}\mathbf{\widehat{n}_{in}}\cdot\nabla I_D(\mathbf{r})
    \label{rad-transfer}
\end{array}
\end{equation}
We derive $j_D^+(\mathbf{r}) $ by solving (\ref{prob-1}), and obtain the boundary conditions for $I'_D,\mathbf{j'}_D$,
\begin{equation}
    I'_D(\mathbf{r}) + \frac{2l}{3}\mathbf{\widehat{n}}\cdot\nabla I'_D(\mathbf{r}) =\frac{2}{v} j_D^+(\mathbf{r}) \mbox{ for any $\mathbf{r}$ at a distance $l$ from the interface}
    \label{prob-2}
\end{equation}
where $\mathbf{\widehat{n}}=-\mathbf{\widehat{n}_{in}}$ is the normal vector of the fictive interface, see Fig.~\ref{fig-bc}.

\medskip

We now derive explicitly the new boundary conditions (\ref{prob-2}) for a slab geometry, considered in the main text.

\subsection{Example: slab geometry}
\label{app:slab geometry}

\begin{figure}
    \centering
    \includegraphics{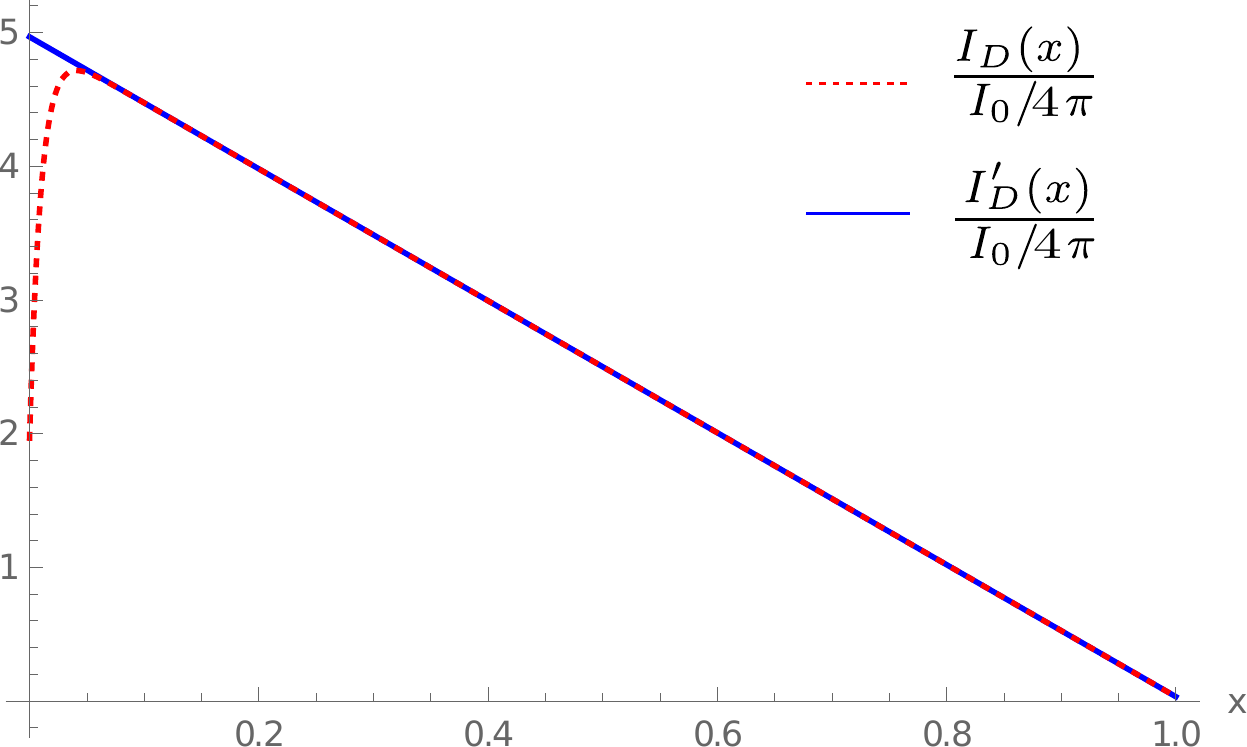}
    \caption{Exact and approximated solutions $I_D$ and $I'_D$ respectively for a slab geometry, for $l/L=0.01$, as functions of the rescaled variable $x\rightarrow x/L$, $x\in[0,1]$. The solutions are normalized by $I_0/4\pi$. }
    \label{fig-plot}
\end{figure}

Consider the case of an infinite slab, of width $L$, illuminated by a homogeneous light beam of intensity $I_0$, see Fig.~\ref{fig-bc}. In this geometry, the radiative transfer equation (\ref{rad-transfer}) becomes
\begin{equation}
\begin{array}{ll}
   I_D(\mathbf{\widehat{s}},x) &= I_D(x) - l\partial_x I_D(x) \\
   \\
  \Rightarrow j_D^+(x) &= \frac{v}{2}I_D(x) - \frac{vl}{3}\partial_x I_D(x)
\end{array}
\label{rel-j}
\end{equation}
In this geometry, the Drude intensity is given by $I_0(\mathbf{r}) = I_0 e^{-x/l}/4\pi$. Solving the exact problem (\ref{prob-1}), we find, in the limit $L\gg l$, 
\begin{equation}
    I_D(l) - \frac{2l}{3}\partial_x I_D(l) = \frac{5I_0}{4\pi}
\end{equation}
We therefore define the boundary conditions to be $j_D^+ = \frac{5I_0}{4\pi}$ at the new boundary (the surface at a distance $l$ from the boundary), which, using eq.(\ref{rel-j}), can be formulated as 
\begin{equation}
    I'_D(\mathbf{r}) + \frac{2l}{3}\mathbf{\widehat{n}}\cdot\nabla I'_D(\mathbf{r}) = 5I_0(\mathbf{r}),
    \label{prob-2-slab}
\end{equation}
where $\mathbf{\widehat{n}}$ is a unit vector normal to the surface, and we recover eq.(2).
Let's now compare the exact and approximated solutions. The approximated solution to (\ref{prob-2-slab}) is 
\begin{equation}
     I'_D(x) = \frac{I_0}{4\pi}\frac{-5+5e^{-L/l}}{4l/3+L}x + \frac{I_0}{4\pi}\frac{5L+\frac{10l}{3}(1+e^{-L/l})}{4l/3+L}  
\label{id-slab}
\end{equation}
In comparison, the exact solution, obtained from (\ref{prob-1}), is
\begin{equation}
    I_{D}(x) = \frac{I_0}{4\pi}\frac{-5+e^{-L/l}}{4l/3+L}x+\frac{I_0}{4\pi}\frac{5L+\frac{2l}{3}(5+e^{-L/l})}{4l/3+L} - \frac{3I_0}{4\pi}e^{-x/l} \, ,
    \label{sol-2}
\end{equation}
hence the two solutions (\ref{id-slab}) and (\ref{sol-2}) differ only by exponentially decreasing terms, see Fig.~\ref{fig-plot}.

\section{Cumulants of the fluctuating radiative force}
\label{app:cumulants of the radiative force}

Let us consider the cummulant generating function (CGF) of  $f = \int d \mathbf{r} \, \mathbf{j}(\mathbf{r}) \cdot \widehat{n} $, namely 
\begin{equation}
    \mu(\lambda) = \log \langle e^{\lambda f} \rangle .
\end{equation}
 The purpose of this section is to show that $\partial_\lambda \mu(0) = \langle f \rangle  = \int d \mathbf{r} \, \mathbf{j}_D(\mathbf{r})\cdot \widehat{n} $ and 
$\partial_{\lambda \lambda} \mu(0) = \int d \mathbf{r} \, \sigma_D(\mathbf{r})$. 

To do so, we first write explicitly the path integral formulation for the cummulant generating function 
\begin{equation}
    e^{\mu(\lambda)} = \int \mathcal{D}I \mathcal{D}j \mathcal{D} p  \exp{ \left( -\int d \mathbf{r} \, \frac{1}{2\sigma} \left(\mathbf{j}+D \nabla I \right)^2 - \lambda \mathbf{j} \cdot \widehat{n} + p \nabla \cdot \mathbf{j} \right) }.   
\end{equation}
The introduction of the $p$  variable -- a Lagrange multiplier --  ensures a divergence free current in the bulk. Integrating the Gaussian integral in $j$, we find \begin{equation}
\label{eq: Hamiltonian of CGF}
    e^{\mu(\lambda)} = \int \mathcal{D}I  \mathcal{D} p \,  e^{ 
    \int d \mathbf{r} \, \mathcal{H}(I,p)(\mathbf{r}) },
\end{equation}
where $\mathcal{H} = -D \nabla I \nabla p + \frac{1}{2}\sigma (\nabla p)^2$ and we redefine $p \rightarrow p+\lambda n $ with $ n \equiv \widehat{n}_x x + \widehat{n}_y y + \widehat{n}_z z   $ and $\widehat{\mathbf{n}}  = (\widehat{n}_x,\widehat{n}_y,\widehat{n}_z)$. Since we are dealing with a weak noise theory, the CGF is dominated by a saddle point solution, given by the saddle equations
\begin{eqnarray}
\label{eq: Hamilton equations}
\frac{\delta \mathcal{H}}{\delta p}  =0 & \Rightarrow &  \nabla \cdot (D\nabla I - \sigma \nabla p  )=0    \\
\frac{\delta\mathcal{H}}{\delta I}  =0 & \Rightarrow &  D\Delta p + \frac{1}{2}\sigma' (\nabla p  )^2=0 ,  
\end{eqnarray}
where $\sigma'(I) = \frac{\delta}{\delta I} \sigma(I)$. The boundary conditions for $I$ is left unchanged as in the main text and such that $p = \lambda n$ on the boundary. Notice that what we have done is simply moving from a Lagrangian picture to a Hamiltonian one. The Hamiltonian picture is more straightforward in this case, where we aim to calculate the first two cumulants of the CGF at $\lambda=0$. A general solution of \eqref{eq: Hamilton equations} is hard to obtain. However, to evaluate $\mu(\lambda)$ to second order in $\lambda$, it is sufficient to consider the perturbative solution 
\begin{eqnarray}
I(\mathbf{r}) &=& I_D(\mathbf{r}) + \lambda \delta I_1 (\mathbf{r}) + O(\lambda^2) \\ \nonumber  
p(\mathbf{r}) &=& \lambda \delta p_1 (\mathbf{r}) + O(\lambda^2).
\label{second-order-sol}
\end{eqnarray}
Solving the saddle equations to first order in $\lambda$ we find 
\begin{eqnarray} 
\label{eq:first order sols}
    D \delta I_1(\mathbf{x}) &=& \widehat{\mathbf{n}} \cdot \int g(\mathbf{x},\mathbf{y}) d\mathbf{y} \nabla_{\mathbf{y} }\sigma_D(\mathbf{y})  \\ \nonumber 
    \nabla \delta p_1 &=& \widehat{\mathbf{n}} , 
\end{eqnarray}
where $\Delta_{\mathbf{x}}g(\mathbf{x},\mathbf{y}) = \delta^d (\mathbf{x}-\mathbf{y})$ defines the Green function of the Laplacian with vanishing boundary conditions. 
Plugging the solutions \eqref{eq:first order sols} into \eqref{eq: Hamiltonian of CGF}, we find to second order in $\lambda$ that indeed  $\langle f \rangle  = \int d^d \mathbf{r} \, \mathbf{j}_D(\mathbf{r})\cdot \widehat{\mathbf{n}} $ and 
$\langle f^2 \rangle_c  = \int d \mathbf{r} \, \sigma_D(\mathbf{r})$.

\section{Dimensionless scaling of the Langevin equation}
\label{app: dimensionless scaling of Langevin}

The purpose of this section is to show that the strength of the noise $\boldsymbol{\xi}$ in the Langevin equation (3),
\begin{equation}
    \mathbf{j} = - D \nabla I(\mathbf{r}) + \boldsymbol{\xi} 
\end{equation}
is, upon proper rescaling, proportional to the dimensionless parameter $1 / g_\mathcal{L} \ll 1$. To that purpose, 
we rescale the spatial coordinates with respect to the length scale $\mathcal{L}$: $ \mathbf{\tilde r} = \mathbf{r}/\mathcal{L} $, $\tilde \nabla  = \nabla_{\mathbf{\tilde r}}= \mathcal{L}\nabla$. Furthermore, we rescale the Langevin equation by dividing by the diffusion constant $D$ and by $I_0$, a typical strength of the external illumination defining $\tilde{I} = I/I_0$. This implies 
\begin{equation}
    \mathbf{\tilde j} = - \tilde\nabla \tilde I(\mathbf{\tilde r}) + \boldsymbol{\tilde \xi }, 
\end{equation}
where $( \tilde{ \mathbf{j}} , \tilde{ \xi} ) \equiv \frac{\mathcal{L}}{  D I_0} (\mathbf{j}, \xi )$. Using the fact that $\delta  (\mathcal{L} \, (\mathbf{\tilde{r}_1 }-\mathbf{\tilde{r}_2 }))=\delta  (\mathbf{\tilde{r}_1 }-\mathbf{\tilde{r}_2 })/\mathcal{L}^3$, we obtain 
\begin{equation}
\langle \tilde{\xi}_\alpha  (\mathbf{\tilde{r}_1 }) \tilde{\xi}_\beta  (\mathbf{\tilde{r}_2 }) \rangle = \frac{2}{g_\mathcal{L}} \tilde{I}^2(\mathbf{\tilde{r}_1 }) \delta  (\mathbf{\tilde{r}_1 }-\mathbf{\tilde{r}_2 }) \delta_{\alpha \beta}.
\end{equation}
Recall that $g_{\mathcal{L}} = \frac{k^2 l^2 }{3\pi } \frac{\mathcal{L}}{l} \gg 1 $ due to the limits taken $k l \gg 1$ and $\mathcal{L}>l$.


\bibliographystyle{unsrt}
 \bibliography{biblio}   


\end{document}